\documentclass[]{svjour3}

\usepackage{graphicx,amsmath,amssymb,enumitem}
\usepackage{mathrsfs,eufrak,color}
\begin{document}


\title{Tensor Network Representation and Entanglement Spreading in Many-Body Localized Systems: A Novel
Approach}

\titlerunning{Tensor Network Representation and Entanglement Spreading . . .}


\author{Z. Gholami \and Z. Noorinejad \and M. Amini \and E. Ghanbari-Adivi$^*$}

\authorrunning{Z. Gholami $et~al$ . . .} 

\institute{Z. Gholami \at
              Faculty of Physics, University of Isfahan, Isfahan   81746-73441, Iran \and
           Z. Noorinejad \at
              Department of Physics,  Islamic Azad University - Shahreza Branch~(IAUSH), Shahreza, Iran \and
           M. Amini \at
              Faculty of Physics, University of Isfahan, Isfahan   81746-73441, Iran\\
              Tel.: +98-31-37934833\and
           E. Ghanbari-Adivi \at
              Faculty of Physics, University of Isfahan, Isfahan   81746-73441, Iran\\
              Tel.: +98-31-37934775\and
              \email{ghanbari@phys.ui.ac.ir}}
\date{Received: date / Accepted: date}
\maketitle
\begin{abstract}
A novel method has been devised to compute the Local Integrals of
Motion~(LIOMs) for a one-dimensional many-body localized system. In
this approach, a class of optimal unitary transformations is deduced
in a tensor-network formalism to diagonalize the Hamiltonian of the
specified system. To construct the tensor network, we utilize the
eigenstates of the subsystems' Hamiltonian to attain the desired
unitary transformations. Subsequently, we optimize the eigenstates
and acquire appropriate unitary localized operators that will
represent the LIOMs tensor network. The efficiency of the method was
assessed and found to be both fast and almost accurate. In framework
of the introduced tensor-network representation,  we examine how the
entanglement spreads along the considered many-body localized system
and evaluate the outcomes of the approximations employed in this
approach. The important and interesting result is that in the
proposed tensor network approximation, if the length of the blocks
is greater than the length of localization, then the entropy growth
will be linear in terms of the logarithmic time. Also, it has been
demonstrated that, the entanglement can be calculated by only
considering two blocks next to each other, if the Hamiltonian has
been diagonalized using the unitary transformation made by the
provided tensor-network representation.
\end{abstract}

\section{Introduction\label{Sec01}}
Many-body localization~(MBL) has a relatively long story. In 1958,
Anderson showed that due to the destructive interference of the
matter waves in a non-interacting system in the presence of an
impurity, the eigenstates can be localized and, despite the
existence of the tunneling probability, the defusing phenomenon can
be lost~\cite{Anderson}. It has been thought that in the presence of
interaction, due to the hierarchy associated to the Fock space of a
many-body system, the related eigenstates are necessarily
non-localized and they satisfy the eigenstate-thermalization
hypothesis (ETH). In 2006, Basko, Aleiner, and Altshuler applied
perturbation theory to show that in the presence of interaction,
there is a possibility of localization in disordered
systems~\cite{Basko}. In 2007, Oganesyan and Huse argued the
existence of the motion constants in a typical many-body system by
examining the spectrum of a Heisenberg XXZ spin-1/2
chain~\cite{Oganesyan}.\par
After the above mentioned pioneering works, the research on the
properties of many-body localized systems or abbreviated MBL systems
and the quantum phase transition between the thermal and MBL phases
was noticed by physicists~\cite{Imbrie01,Pal01}.  One of the
important concepts in MBL problems is the existence of motion
constants or the Local Integrals of
Motion~(LIOMs)~(See~Ref.~\cite{Imbrie01} and references therein).
Through a number of studies on typical MBL systems, it has been
shown for these systems that usually an extendable number of the
localized operators can be found that obey the Pauli matrix algebra
and commute with the Hamiltonian of the system, these operators are
called LIOMs. Many of the behaviors exhibited by such a system can
be explained using the LIOMs picture. For example, the LIOMs picture
can be utilized to explain the logarithmic growth of the
entanglement entropy in time.\par
The importance of the motion constants of the MBL systems has led to
many efforts to obtain an explicit form of the LIOMs for the XXZ
models~\cite{Imbrie01,Pal01,Luitz01,Goihl01,Serbyn01,Peng01,Adami01,Gholami01,Huse01,Ros01,Imbrie02,Chandran01,Geraedts01,Rademaker01,OBrien01,Kulshreshtha01,Pekker01,Chandran02,Pollmann01,Wahl01}.
In some of these efforts, it has been attempted to calculate LIOMs
by the exact digitalization of the Hamiltonian of the considered
system. For example,
in~\cite{Goihl01,Serbyn01,Peng01,Adami01,Gholami01}, the appropriate
unitary operators for calculation of the LIOMs has been obtained by
using the exact diagonalization of the Hamiltonian and ordering of
the energy eigenstates of the system.  Using the obtained results
the localization length as well as the localization aspects of the
specified system have been investigated.\par
The construction of LIOMs based on the exact diagonalization leads
to interesting theoretical results. However, the computations
related to this method are very time-consuming and uneconomical, so
that with today's very fast computers, at most, the Hamiltonian of
the spin space with length 24 can be exactly diagonalized. Due to
this limitation, this method is not applicable for the system with
larger length and it is natural to look for some faster methods to
diagonalize the Hamiltonian and construct the LIOMs of the MBL
systems. One of the extendable methods proposed based on the
localization behavior of the Hamiltonian is the use of the unitary
transformations in the framework of the tensor
networks~\cite{Pollmann01,Wahl01}. In this regard, it was proved for
the first time that by successively multiplying the unitary matrices
of the two-spin space, which are multiplied together in tensor form,
a unitary matrix can be created that diagonalizes the Hamiltonian in
a very good approximation and leads to LIOMs whose commutation
magnitude with the Hamiltonian is very small~\cite{Pollmann01}. In
another study~\cite{Wahl01}, it is shown that it is possible to
increase the length of the spin space of unitary operators and
instead make a unitary operator from the successive matrix
production of two corresponding matrices of the unitary
operators.\par
The tensor network is extendable, so the length of the MBL system
can be easily increased in such a way that the computations increase
linearly rather than powers of 2. However, because in
studies~\cite{Pollmann01} and~\cite{Wahl01}, it has been tried to
optimize the unitaries by minimizing the magnitude of the commutator
of the approximate integrals of motion and the Hamiltonian, this
method does not work efficiently and does not reduce the difficulty
of solving the problem. The reason for this fact is that by
increasing the length of the chain, there are many continuous
parameters that need to be optimized. The optimization process is
time-consuming and therefore the problem of uneconomical
computations still remains in the introduced procedure based on the
tensor networks.\par
In our previous study, a multi-body expansion of LIOM creation
operators of the system in the MBL regime has been explicitly
performed and the associated coefficients have been obtained in
terms of different number of pairs of
quasi-particles~\cite{Gholami01}. In the present study, we will
introduce a simple and optimal method to obtain a sequence of the
local unitary matrices to construct the unitary transformation
represented as a tensor network diagonalizing the entire
Hamiltonian. In this method, instead of focusing on the optimization
of the continuous parameters, we focus on diagonalizing the
Hamiltonians of the subspaces and ordering their eigenspectrums
using the optimization procedure introduced in Ref.~\cite{Wahl01}.
In the optimaization procedure, the magnitude of the commutator of
the obtained approximate integrals of motion and the Hamiltonian has
been minimized which it is usually time-consuming. In order to
examine the accuracy of the introduced method, the upper limit that
the local operator can construct a LIOM in a system with an infinite
length is obtained, and based on that, the accuracy of the tensor
network method with a given block length is indicated.\par
Entropy growth in MBL systems is one of the fundamental differences
between the behavior of these types of systems and thoes systems
that obey ETH, so it has been widely investigated in the
literature~\cite{Kulshreshtha02,Znidaric01,Bardarson01,Serbyn02,Iemini01,Dumitrescu01,Znidaric02,Chiaro01,Nanduri01}.
It is surprising to investigate the spread of entanglement in an MBL
system using the introduced tensor-network representation as an
application of the proposed approach. As mentioned above, the
unitary operator obtained from the tensor network diagonalizes the
entire Hamiltonian with a good approximation. Consequently, after
applying the unitary operator on the Hamiltonian, the non-diagonal
elements can be neglected. In these calculations, it has been shown
that in the tensor network approach to compute the entropy growth,
only the boundary blocks are involved. The important and interesting
result obtained is that for the blocks of small lengths, the growth
of entanglement quickly saturates to a constant value. But if the
length of the blocks is chosen to be large and comparable to the
localization length of the LIOMs, the entropy growth will be linear
in terms of the logarithm of time. This result is interesting and
important because it is in complete agreement with what was obtained
in Ref.~\cite{Bardarson01} using the exact diagonalization of the
entire Hamiltonian for the MBL systems. In fact, this consistency
indicates that if the length of the blocks is chosen large enough,
the approximations used in the present tensor network approach will
be very useful and efficient.\par
Another important and intersting result which we have achieved is
that when the unitary transformation provided by the current
tensor-network representation is used to diagonalize the
Hamiltonian, considering only two blocks next to each other is
sufficient to compute the entanglement.\par
The plan of the rest of the paper is as follows. The theoretical
framework of the proposed tensor-network approach for calculation
the LIOMs of the considered MBL system is briefly described in the
next section. By presenting some examples of the LIOMs calculations,
the accuracy of the method has also been discussed in this
section.The method is applied to deal with the entanglement entropy
spreading, and the related results and discussions are given in the
third section. Finally, the conclusion remarks are given in the last
section.\par
\begin{figure}[t]
\begin{center}
 \center{\includegraphics[width=0.5\linewidth]{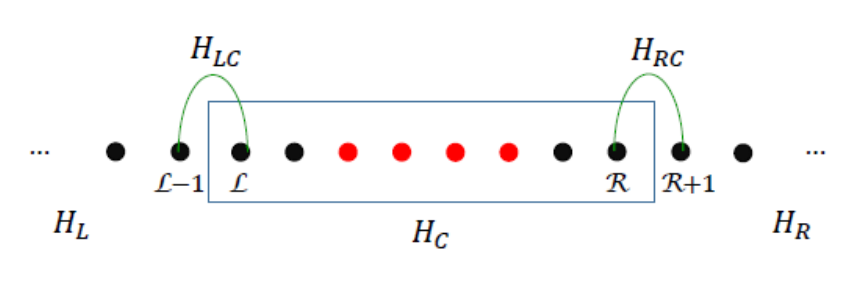}} 
\end{center}
\vspace*{-3mm}   \caption{Segmentation of the Hamiltonian of a spin
chain in five zones. $H_{C}$ is the Hamiltonian of the central
domain, $H_{L}$ and $H_{R}$ the the Hamiltonians of the left-hand
and right-hand sides,respectively. Moreover, $H_{LC}$ and $H_{RC}$
are the interacting Hamiltonians between the left and right sides
with the central part, respectively, which both just defined on the
boundaries of the left and right.\label{Fig01}}
\end{figure}
\section{Calculation of the approximate LIOMs\label{Sec02}}
The main purpose of this research is to demonstrate how to compute
the LIOMs for a one-dimensional~(1D) spin-1/2 chain with an
arbitrary length. Additionally, we will explore how these calculated
LIOMs allow us to examine the dynamics of the entanglement in such
interacting systems.\par
\subsection{Model and method\label{Sec02Sub01}}
Let us consider a Heisenberg XXZ spin-1/2 chain which has a length
of N, is subjected to open boundary conditions and is influenced by
a random static magnetic field in the z-direction. Hamiltonian of
this specified system can be written as
\begin{equation}\label{Eq01}
H = J \sum_{i=1}^{N-1} (\sigma_i^x \sigma_{i+1}^x + \sigma_i^y
\sigma_{i+1}^y) + \Delta \sum_{i=1}^{N-1} \sigma_i^z \sigma_{i+1}^z
+ \sum_{i=1}^N h_i \sigma_i^z,
\end{equation}
where $\sigma_i^{x,y,z}$ denote the Pauli operators acting on the
$i$th spin of the chain and independent random magnetic fields $h_i$
are derived uniformly from a distribution $[-W,W]$, where $W$ is
called disorder strength. We fix exchange interaction coupling at
$J=1$ and the parameter $\Delta$ indicates the strength of the
many-body interaction term. This standard model of MBL is known to
exhibit a phase transition at a critical disorder strength $W =
W_c\sim 7$ from an ergodic phase to an MBL one
\cite{Pal01,Luitz01,Goihl01} and on behalf of this transition,
thermalization fails.
As previously discussed in some other studies, for $W>7$,
outstanding feature of the MBL regime is the existence of LIOMs
\cite{Serbyn01,Huse01}, $\tau_i^z$, which satisfy the following
properties \cite{Ros01,Imbrie02}:
\begin{enumerate}[label=(\roman*)]
\item they need to obey Pauli spin algebra commutator relations.  \label{Item01}
\item they are conserved and independent operators.  \label{Item02}
\item they are exponentially (quasi-)local operators which reads:
\begin{equation}\label{Eq02}
\Vert [ \tau_{i}^{z}, \sigma_{j}^{\alpha} ] \propto e^{-\vert i-j
\vert / \xi},
\end{equation}
   \label{Item03}
\end{enumerate}
where $\alpha = x,y,z$ and $\xi$ is the localization length of the
LIOMs.\par
Regarding to the property~\ref{Item02}, by definition, $\tau_i^z$
commute with the Hamiltonian in such a way that an effective
phenomenological Hamiltonian can be written with respect to them as
\cite{Pal01,Chandran01}
\begin{equation}\label{Eq03}
H=\sum_i \varepsilon_i \tau_i^z + \sum_{ij} J_{ij} \tau_i^z \tau_j^z
+ \sum_{ijk} J_{ijk} \tau_i^z \tau_j^z \tau_k^z + \cdots,
\end{equation}
in which $J_{ij}$ vanishes by increase the distance of $|i-j|$
\cite{Goihl01,Chandran01} and deep in the MBL phase, due to the fact
that the localization length of LIOMs is small, the higher-order
terms in Eq.~\eqref{Eq03} will be negligible. Furthermore, it is
easily inferred that these localized pseudo-spin operators
$\tau_i^z$ can be related to their physical spins $\sigma_i^z$ using
a local unitary transformation. To put it another way, we have
\begin{equation}
\tau_i^z=U \sigma_i^z U^{\dag}. \label{Eq04}
\end{equation}
To determine a complete set of LIOMs for a finite-size system (where
numerical study is directly available) which satisfy the previously
mentioned properties~\ref{Item01}-\ref{Item03}, a great number of
various methods including labeling the eigenstates of the system by
their corresponding LIOM-eigenvalues uniquely
\cite{Serbyn01,Huse01}, computing an infinite-time average of
initially local operators~\cite{Goihl01,Chandran01,Geraedts01} and
exact diagonalization
techniques~\cite{Peng01,Rademaker01,OBrien01,Kulshreshtha01,Adami01}
have been presented among which the non-perturbative fast and
efficient scheme developed in Ref.~\cite{Adami01} is our matter of
interest. In the mentioned research, it has been suggested that the
intended algorithm can be implemented via rearranging an optimized
set of eigenstates of the system in a quasi-local unitary operator
explicitly which maps the physical spin operators onto their
effective spins operators. Such an ordered set of the eigenbasis can
be achieved by allocating an integer index number to each eigenstate
which identifies its order in the desired set. This index number can
be recognized by locating the original basis vector of the
Hilbert-space on which that eigenstate has the largest absolute
amplitude among all the eigenstates of the system. To the best of
our knowledge, this scheme provides the well rearranged $U$ required
for the construction of pseudo-spin operators.\par
\begin{figure}[t]
\begin{center}
\includegraphics[width=9cm]{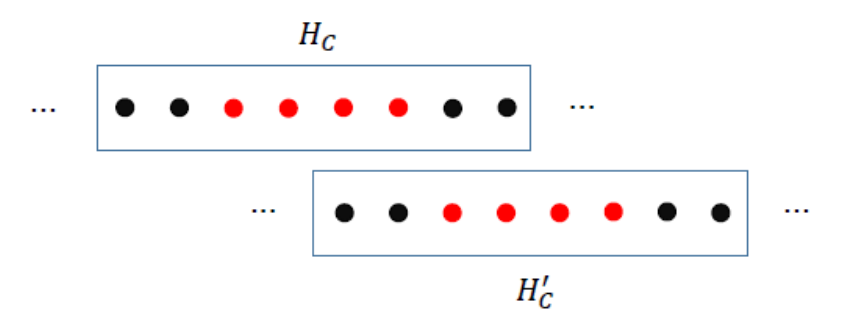}
\end{center}
\vspace*{-3mm}   \caption{We assume that $H_C$ is the Hamiltonian
made on $l$ sites by diagonalization of which we will be able to
construct $l\over 2$ LIOMs of the middle of the subsystem (top red
dots). Now for calculation of the next ${l\over 2}$ LIOMs of the
chain (bottom red dots), we ought to shift the subsystem as
demonstrated in the figure. This leads to another Hamiltonian $H'_C$
which does not commute with $H_C$.\label{Fig02}}
\end{figure}
Although such studies on finite-size systems has created a deep
insight into the MBL phase, practically we encounter with the
systems of infinite length at the thermodynamic limit. So conducting
a research which targets construction of LIOMs on an arbitrary long
chain will made a great progress in this realm. To this end, as it
is previously considered in some various
studies~\cite{Pekker01,Chandran02,Pollmann01,Wahl01}, one can
utilize the localization of LIOMs in order to investigate a spin
chain of infinite length in the MBL phase. It means that, instead of
finding a unique unitary transformation for the entire system by
diagonalization of the Hamiltonian in the total Hilbert space, LIOMs
of the problem can be computed by means of tensor networks
\cite{Pekker01,Chandran02,Pollmann01,Wahl01}. One of the suggested
strategies, which has been presented in Ref.~\cite{Kulshreshtha02},
introduces an algorithm, labeled the inchworm algorithm to calculate
the expectation value of a product of local operators on a large
system while working in a Hilbert space of computationally
manageable size. We initialize the process using a subsystem on the
rightmost window of length~$l$, then expand the system leftward and
to keep the working space manageable, they subsequently contract the
system from the right. Inspired by this method, we divide the
Hamiltonian into five zones shown in Fig.~\ref{Fig01}, So, the
Hamiltonian can be written as
\begin{equation}
H=H_C + H_L + H_R + H_{LC} + H_{RC}, \label{Eq05}
\end{equation}
where $H_C$ is the Hamiltonian of a region called central domain of
length $l$ in which $\tau_i^z$ is decided to be calculated. In other
words, if $\tau_i^z$ is the pseudo-spin defined on a site of the
interval $[n_0-{l\over 4},n_0+{l\over 4}]$, in which $n_0$ is the
middle of the central domain, $H_C$ is the Hamiltonian belonging to
$[n_0-{l\over 2},n_0 + {l\over2}]$. $H_L$ is the Hamiltonian of the
rest of the system in left side and $H_R$ is that of the right side.
$H_{LC}$ and $H_{RC}$ are the interacting Hamiltonians between the
left and right sides with the central part, respectively. These
interaction Hamiltonians can be written in their explicit forms as
\begin{equation}\label{Eq06}
H_{RC} = J(\sigma_{\mathcal{R}}^x \sigma_{\mathcal{R}+1}^x +
\sigma_{\mathcal{R}}^y \sigma_{\mathcal{R}+1}^y) + \Delta
\sigma_{\mathcal{R}}^z \sigma_{\mathcal{R}+1}^z,
\end{equation}
and
\begin{equation}\label{Eq07}
H_{LC}=J(\sigma_{\mathcal{L}-1}^x \sigma_{\mathcal{L}}^x +
\sigma_{\mathcal{L}-1}^y \sigma_{\mathcal{L}}^y) + \Delta
\sigma_{\mathcal{L}-1}^z \sigma_{\mathcal{L}}^z,
\end{equation}
in which $\sigma_m^{x,y,z}$ act on site $m \in \lbrace
\mathcal{L}-1, \mathcal{L} , \mathcal{R} , \mathcal{R}+1 \rbrace$ as
shown in Fig.~\ref{Fig01}. The essential point is that the
Hamiltonians of $H_R$ and $H_L$ have no contribution in construction
the desired pseudo-spin. Hence, for calculating $\tau_i^z$, merely
the consideration of $H_C$ will be sufficient and we are able to
find exact LIOMs by using the scheme proposed in Ref.~\cite{Adami01}
for $H_C$ which reads
\begin{equation}\label{Eq08}
[\tau_i^z,H_C]=0.
\end{equation}
It should be noted that although diagonalization of the total
Hamiltonian $H_C$ and obtaining $\tau_i^z$ belonging to this $H_C$
leads to the exact LIOMs, its generalization to the whole of the
system violates the independence of pseudo-spins $\tau_i^z$. This
happens due to the fact that if we are going to compute
$(n_0+{l\over 2}+1)$th LIOM according to this method, we should
utilize $H'_C$ instead of $H_C$ and as the Fig.~\ref{Fig02} shows
obviously, these two Hamiltonians do not commute.
\begin{equation}\label{Eq09}
[H_{C}^{\prime},H_{C}]\neq 0.
\end{equation}
Therefore, the condition of orthogonality and independence of
pseudo-spins $\tau_i^z$ and $\tau_{i'}^z$ will not be held as well
since they clearly do not satisfy the commutation relation. However,
if the central-domain length is considered such that it is much
greater than the localization length of LIOMs, the commutator
magnitude will be very small and as a result, it can be completely
neglected.\par
The other studious works which have been done to calculate such
approximate LIOMs are Refs.~\cite{Pollmann01,Wahl01} in which the
authors have exploited the locality of integrals of motion and
introduced tensor networks techniques by which such numerous local
unitary transformations can be obtained. In Ref.~\cite{Pollmann01},
a two-leg four-layer tensor network, using the variance of the
energy summed over all approximate many-body eigenstates as their
figure of merit, have been considered and in Ref.~\cite{Wahl01}, the
authors have made an effort to keep the layers in two by increase
the number of legs and they have optimized the unitaries by
minimizing the magnitude of the commutator of the approximate
integrals of motion and the Hamiltonian, which can be done in a
local manner. Although these mentioned studies have presented some
very insightful analyses, the milestone is that minimization of
their figure of merits directly to find the optimum unitaries impose
a great deal of computational cost due to ascending number of
parameters. Thus, implementing the procedure will be impossible for
the systems of arbitrary length by increase the number of legs.\par
Equipped with what expressed above, let us now explain our noble and
efficient scheme. Actually, we tend to acquire $\tau_i^z$ such that
this LIOM satisfies the following relation,
\begin{equation}\label{Eq10}
[\tau_i^z,\sigma_j^\alpha]=0, \qquad\ \text{for} \quad |j-i|>{l\over
2}.
\end{equation}
In other words, we are looking for the LIOMs arisen from a unitary
transformation which acts only in the central domain. The crucial
point is that these obtained LIOMs need to be independent which
means,
\begin{equation}\label{Eq11}
[\tau_i^\alpha,\tau_j^\beta]= 0, \qquad\ \text{if} \quad i\neq j.
\end{equation}
To this end, inspired by the tensor networks method proposed in
Ref.~\cite{Wahl01} used for the blocks of an arbitrary number of
legs, we keep the number of layers in two and choose the length of
the subsystem, on which $H_C$ is erected, 'even' that we denote it
by $l$.
\begin{figure}[t]
\begin{center}
\includegraphics[width=14cm]{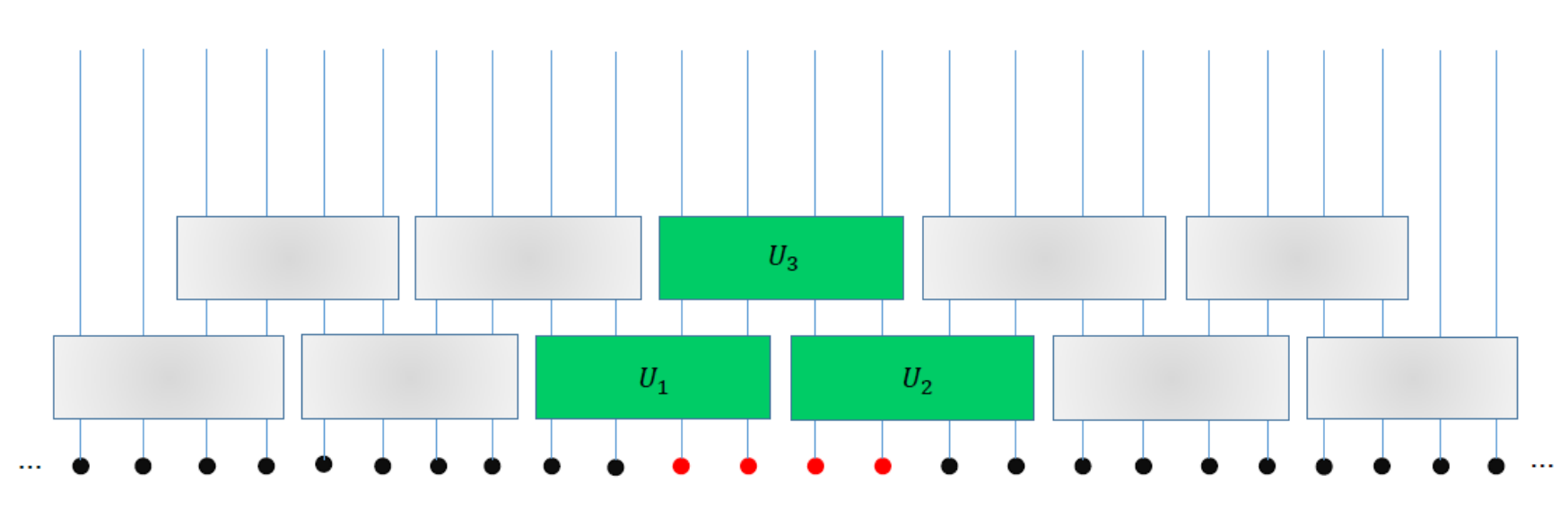}
\end{center}
\vspace*{-3mm} \caption{This diagram exhibits a two-layer tensor
network of typical four-leg unitary operators applied for finding
$\tau_i^z$ on red sites. To this end, independent of how long the
chain is, $U_1$ , $U_2$ and $U_3$ will suffice to calculate our
desired LIOMs on red sites. Although, in this diagram, we choose the
number of legs four to illustrate the gist of our procedure, any
other even number of legs can be chosen.\label{Fig03}}
\end{figure}
For the first layer, we divide this subsystem into two ${l\over
2}$-leg blocks as half cut of the subsystem shown in
Fig.~\ref{Fig03}. These blocks have the Hamiltonians of form of
\begin{equation}
H_1 = J \sum_{i=1}^{{l\over 2}-1} (\sigma_i^x \sigma_{i+1}^x +
\sigma_i^y \sigma_{i+1}^y) + \Delta \sum_{i=1}^{{l\over 2}-1}
\sigma_i^z \sigma_{i+1}^z + \sum_{i=1}^{l\over 2} h_{i}\sigma_i^z,
\label{Eq12}
\end{equation}
and
\begin{equation}
H_2=J \sum_{i={l\over 2}+1}^{l-1} (\sigma_i^x \sigma_{i+1}^x +
\sigma_i^y \sigma_{i+1}^y) + \Delta \sum_{i={l\over 2}+1}^{l-1}
\sigma_i^z \sigma_{i+1}^z + \sum_{i={l\over 2}+1}^l h_i\sigma_i^z,
\label{Eq13}
\end{equation}
where $H_1$ is the Hamiltonian of the left block and $H_2$ is that
of the right one. By employing exact diagonalization, we will
achieve two local unitary transformations. It is noticeable that to
this end, we use the same method presented in Ref.~\cite{Adami01}
for finding the best permutation of the eigenstates to access the
appropriate unitaries. Using such an approach to deal with the
problem will intensively reduce the computational cost rather than
the method utilized in Refs.~\cite{Pollmann01} and~\cite{Wahl01} and
enable us to generalize our computations for the blocks on which
exact diagonalization is numerically available. Thus, we have:
\begin{equation}
(H_{i})_{diag}=U_{i}^{\dagger}H_{i}U_{i}, \qquad \text{for} \qquad
i=1,\ 2. \label{Eq14}
\end{equation}
That being said, in order to obtain the Hamiltonian of the second
layer let us complete our calculations as follows:
\begin{equation}
H_{3}=H_{01}+H_{12}+H_{02}, \label{Eq15}
\end{equation}
where $H_3$ is the Hamiltonian of the second layer defined on a
$l\over 2$-leg block in the middle of our determined subsystem. In
this relation, to attain $H_{01}$ and $H_{02}$, we need to project
$(H_1)_{diag}$ on the basis corresponding to the sites of $[{l\over
4},{l\over 2}]$ and similarly $(H_2)_{diag}$ will be projected on
the basis related to the sites of $[{l\over 2}+1,{l\over 2}+{l\over
4}]$ for which we use the partial trace calculation techniques.
Subsequently, expanding the Hilbert space upon the block of
$[{l\over 4},{l\over 2}+{l\over 4}]$ will be done for both $H_{01}$
and $H_{02}$. Eventually, by writing $H_{12}$ in our new basis we
will have an access to our desired correct $H_3$. This goal is
feasible as follows:
\begin{equation}
H_{12}=J (\tilde{\sigma}_{l\over2}^x \tilde{\sigma}_{{l\over2}+1}^x
+ \tilde{\sigma}_{l\over2}^y \tilde{\sigma}_{{l\over2}+1}^y)
+\Delta( \tilde{\sigma}_{l\over2}^z \tilde{\sigma}_{{l\over2}+1}^z),
\label{Eq16}
\end{equation}
in which:
\begin{equation}
\tilde{\sigma}_{l\over2}^\alpha = U_1^\dagger \sigma_{l\over
2}^\alpha U_1, \qquad \text{for} \qquad \alpha = x,\ y,\ z. \label{Eq17}
\end{equation}
As pointed out before, the operator of
$\tilde{\sigma}_{l\over2}^\alpha$ also has to be projected on the
basis of $[{l\over4},{l\over2}]$. Analogously, we will do the same
procedure for $\tilde{\sigma}_{{l\over 2}+1}^\alpha$ and project it
on the basis of $[{l\over 2}+1,{l\over 2}+{l\over4}]$. Ultimately,
by expansion of the Hilbert space of those over the basis
corresponding to the block of $[{l\over 4},{l\over 2}+{l\over 4}]$,
these operators will be applicable to create the Hamiltonian of
$H_{12}$. Then by diagonalization of $H_3$, one can achieve the
unitary matrix related to the second layer and subsequently, by
product of individual obtained unitaries we will be able to
calculate a unique unitary for the entire $H_C$ approximately.
\begin{figure}[t]
\begin{center}
\includegraphics[width=9cm]{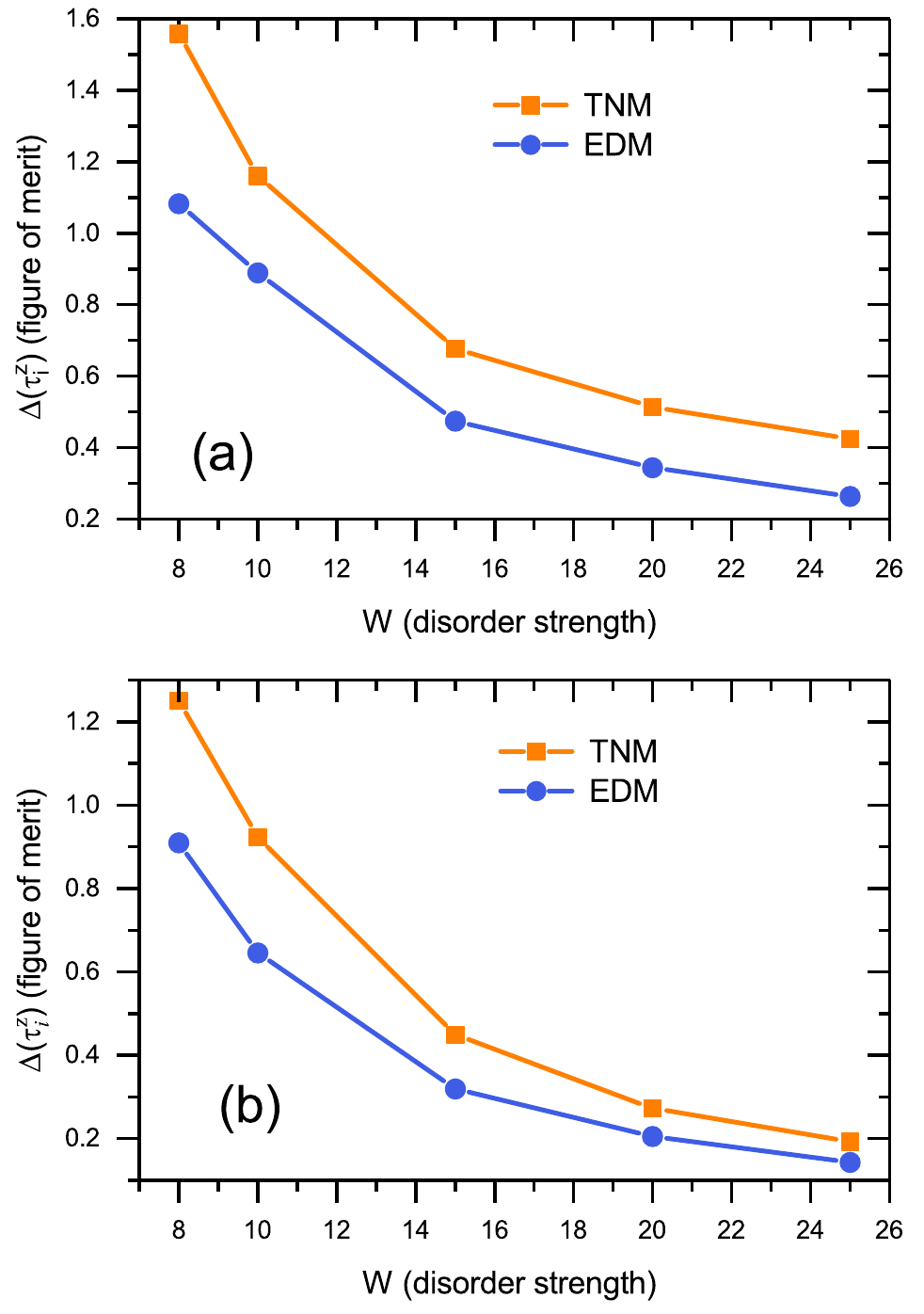}
\end{center}
\vspace*{-3mm} \caption{Using the proposed tensor network
method~(TNM), the value of figure of merit, $\Delta(\tau_i^z)$, has
been calculated in terms of the disorder strength, $W$,  and
compared with the results of the exact diagonalization method~(EDM):
a) TNM~results with block length of~4 correspond to EDM~results for
a chain of~$5$ sites, and b) TNM~results with block length~$6$
correspond to EDM~results for a chain of~$7$ sites. From a physical
point of view, we expect the values compared in each panel to be the
same. The number of realizations for TNM is considered equal to~500,
while it has been taken equal to~$1000$ for the EDM. \label{Fig04}}
\end{figure}
\subsection{Efficacy of the method: how precise the approximate LIOMs are?
\label{Sec02Sub02}}
Having said that, it is inferred that $\tau_i^z$ can be attained
using a local unitary transformation. This unitary transformation can be written in an explicit form of
\begin{equation}
\tau_i^z = U_{Loc} \sigma_i^z U_{Loc}^{\dag}, \label{Eq18}
\end{equation}
where the unitary operator of $U_{Loc}$ only acts on the basis of
central domain. Note that, if $l\gg\xi$ (where $\xi$ is the
localization length of $\tau_i^z$), $\tau_i^z$ will be exact.
However, in general case, if $l$ is of the same order of $\xi$,
$\tau_i^z$ is not accurate and therefore, we ought to introduce a
criterion for investigating the precision of the pseudo-spin
$\tau_i^z$. In this work, we use the figure of merit defined in Ref.~\cite{Wahl01} which reads
\begin{equation}\label{Eq19}
\Delta(\tau_i^z)={1\over 2^{N}}\left({1\over 2}
Tr([\tau_i^z,H][\tau_i^z,H]^\dag)\right)= {1\over
2^N}\left(Tr(H^2)-Tr(H\tau_i^z H\tau_i^z)\right).
\end{equation}
Before continuing the discussion, let us obtain the figure of merit, given in
Eq.~\eqref{Eq19}, for the physical spins themselves. Using
Eq.~\eqref{Eq01}, it is a simple practice to show that
\begin{equation}\label{Eq20}
Tr(H^2)=2^N\left(\sum_ih_i^2 +(N-1)(2J^2+\Delta^2) \right),
\end{equation}
and
\begin{equation} \label{Eq21}
Tr(H\sigma_i^z H \sigma_i^z) = Tr(H'\sigma_i^z H' \sigma_i^z) +
Tr(H_{int}\sigma_i^z H_{int}\sigma_i^z),
\end{equation}
where $H_{int}$ is the term of the Hamiltonian which does not
commute with $\sigma_i^z$. It is easily indicated that
\begin{equation}\label{Eq22}
H_{int} = J (\sigma_{i-1}^x \sigma_i^x + \sigma_{i-1}^y \sigma_i^y +
\sigma_i^x \sigma_{i+1}^x + \sigma_i^y \sigma_{i+1}^y),
\end{equation}
and $ H'=H-H_{int}$. Hence,
\begin{equation}\label{Eq23}
\Delta(\sigma_i^z)= {1\over 2^N} \left( Tr(H^2)-Tr(H\sigma_i^z
H\sigma_i^z) \right) = 8 J^2,
\end{equation}
and in the boundaries, this value is equal to $4 J^2$. Now, we
address the main question of this section. Indeed, we would like to
calculate $\Delta(\tau_i^z)$ generally. To this end, using
Eqs.~\eqref{Eq05}-~\eqref{Eq07}, we have
\begin{equation}\label{Eq24}
\Delta(\tau_i^z)=\Delta_1(\tau_i^z) + \Delta_2(\tau_i^z),
\end{equation}
where $\Delta_1(\tau_i^z)$ is the contribution of $H_C$ and
$\Delta_2(\tau_i^z)$ is the participation of $H_{LC}$ and $H_{RC}$
altogether. It can be easily shown that
\begin{equation}\label{Eq25}
\Delta_1(\tau_i^z)={1\over 2^l}\left( Tr(H_C^2)-Tr(H_C\tau_i^z H_C
\tau_i^z)\right)
\end{equation}
and
\begin{align}\label{Eq26}
\Delta_2(\tau_i^z) & = 2 (2 J^2 + \Delta^2)\notag\\ & - {1\over
2^l}\left[ J^2
Tr(\sigma_\mathcal{L}^x\tau_i^z\sigma_\mathcal{L}^x\tau_i^z)+ J^2
Tr(\sigma_\mathcal{L}^y \tau_i^z\sigma_\mathcal{L}^y \tau_i^z)+
\Delta^2 Tr(\sigma_\mathcal{L}^z \tau_i^z\sigma_\mathcal{L}^z \tau_i^z)\right]\\
& - {1\over 2^l} \left[ J^2 Tr(\sigma_\mathcal{R} ^x \tau_i^z
\sigma_\mathcal{R}^x\tau_i^z)+ J^2 Tr(\sigma_\mathcal{R}^y \tau_i^z
\sigma_\mathcal{R}^y \tau_i^z)+ \Delta^2 Tr(\sigma_\mathcal{R}^z
\tau_i^z \sigma_\mathcal{R}^z \tau_i^z) \right].\notag
\end{align}
 A point of note is that the Hamiltonians of $H_R$ and $H_L$ have no participation in
$\Delta(\tau_i^z)$. Thus, to obtain $\tau_i^z$, including $H_C$
leads to the appropriate result.
Now, in order to gain a limit for $\Delta(\tau_i^z)$, we suppose to
look for a $U$ for the central domain such that it minimizes
$\Delta(\tau_i^z)$ over the region of
$[n_0-{l\over2},n_0+{l\over2}]$. In other words, we use a one-layer
$U$ instead of two-layer unitary matrices. So, according to
Eq.~\eqref{Eq08}, we have $\Delta_1(\tau_i^z)=0$ and owing to the
localization of $\tau_i^z$ the amount of $\Delta_2(\tau_i^z)$ has
definitely a small value. We emphasize that diagonalization of the
total Hamiltonian $H_C$ and finding exact pseudo-spins $\tau_i^z$
corresponding to this $H_C$ leads to the LIOMs for which
$\Delta_0(\tau_i^z)$ is significantly small and it can be be taken
into account as a limit for $\Delta(\tau_i^z)$. In other words, by
means of the scheme indicated in Fig.~\ref{Fig03} we are not able to
make the amount of $\Delta(\tau_i^z)$ less than that one of
$\Delta_0(\tau_i^z)$.\par
To answer the question that ``how accurate are the results obtained
from the proposed tensor-network approximation?", one can calculate
the values of the figure of merit, $\Delta(\tau_i^z)$, by using this
method and compare the obtained results with those corresponding
values obtained from the exact diagonalization of the entire
Hamiltonian. In figure~\ref{Fig04}, we have made a comparison
between the numerical results obtained from the proposed tensor
network representation with block lengths of 4 and 6 and those
related to the exact diagonalization of a part of the Hamiltonian.
From the physics of the problem, we expect that the results
calculated employing the tensor network representation with a block
length of~$4$ are proportional to those obtained using the exact
diagonalization for a chain of~$5$ sites, and those for a block
length of~$6$ are proportional to the exact diagonalization for a
chain of~$7$ sites. As can be seen from the figure, the comparison
shows that the results are completely consistent with our physical
expectation. In other words, for obtaining the appropriate results
of the Hamiltonian diagonalization by using the tensor network
method presented here, one should consider the length of the blocks
equal to the localization length of LIOMs of the specified system.
This comparison emphasizes that by moving away from the phase
transition point, it is possible to achieve the desired results by
considering a normal length for the blocks in the tensor network.
Another noteworthy point in figure~\ref{Fig04} is that figure of
merit, $\Delta(\tau_i^z)$, decreases exponentially with increasing
the disorder strength, W.\par
\begin{figure}[t]
\begin{center}
\includegraphics[width=16cm]{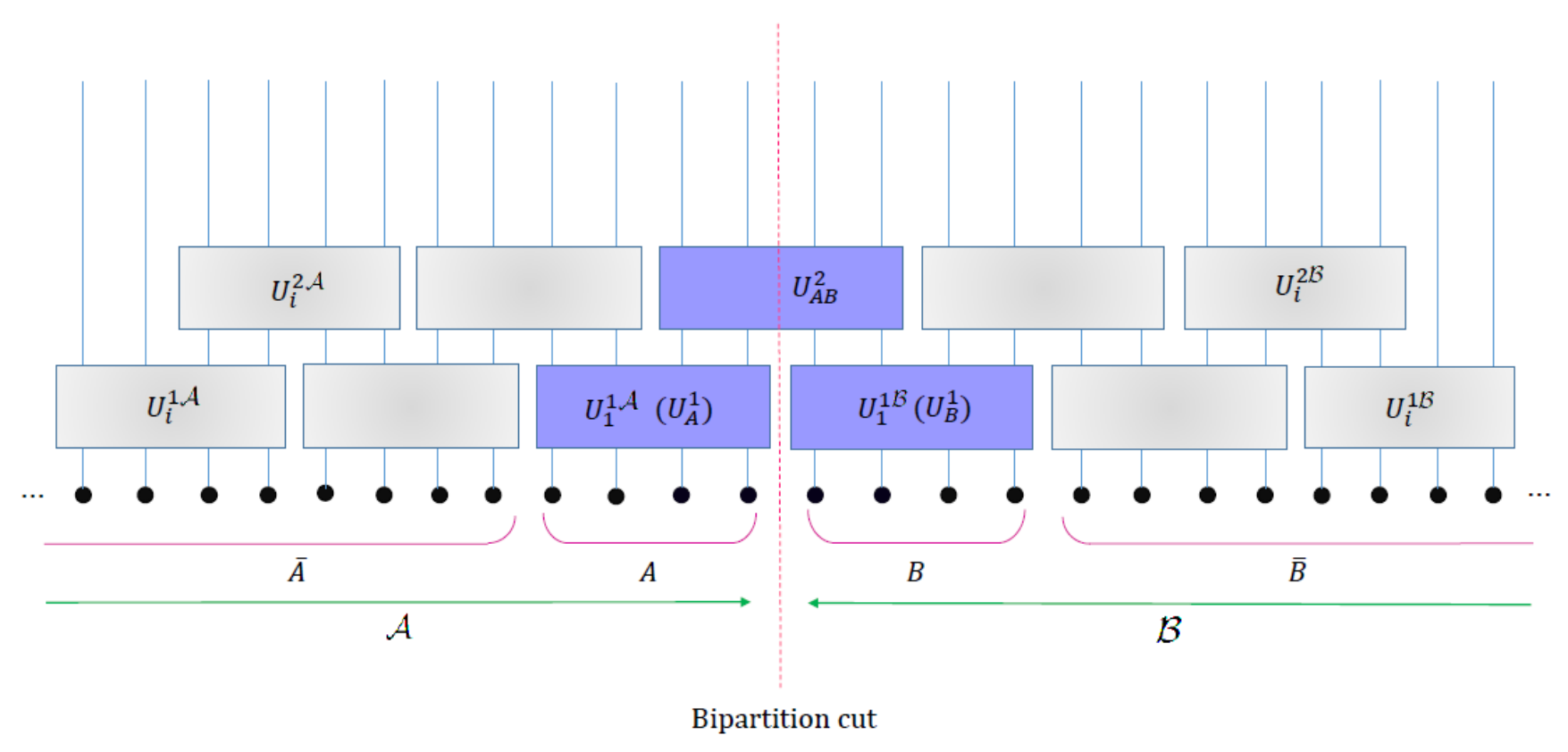}
\end{center}
\vspace*{-3mm} \caption{Schematic representation of the bipartite
entanglement entropy calculation using a two-layer tensor network of
an arbitrary number of legs.\label{Fig05}}
\end{figure}
\section{Calculation of the entanglement generation
rate\label{Sec03}}
One of the most outstanding features of the MBL systems is the
existence of particular behaviour in the entanglement entropy
generation rate in a bipartite system. Recently some precious
studies have shed light on this characteristic of the MBL
regime~\cite{Znidaric01,Bardarson01,Serbyn02,Iemini01,Dumitrescu01}.
Ref.~\cite{Znidaric02} has studied entanglement dynamics in a
diagonal dephasing model in which the intensity of interaction
decays exponentially with distance and has calculated the exact
expression for entanglement growth with time, obtaining in addition
to a logarithmic growth, a sublogarithmic correction. Using the
l-bit picture, correctly describes the MBL phase, implies that the
entanglement in such systems does not grow (just) as a logarithm of
time, as believed so far. Moreover, Chiaro and his co-workers~Ref.~\cite{Chiaro01},
using phase sensitive measurements, have
experimentally characterized logarithmic entanglement growth of the
MBL phase in such systems to determine the spatial and temporal
growth of entanglement between the localized sites. In addition,
they have studied the preservation of entanglement in the MBL phase.
In this work, we are also going to demonstrate that our
above-mentioned approach, presented in Sec.~\ref{Sec02} to compute
LIOMs, can be employed to calculate the entanglement generation rate
and enables us to implement less numerical computations in a
possibly short time. Generally, to investigate the generation of
entanglement, inspired by some ever-done studies, we consider the
initial state as
\begin{equation} \label{Eq27}
\vert \psi \rangle = \vert \psi_\mathcal{A} \rangle \otimes \vert
\psi_\mathcal{B} \rangle,
\end{equation}
in which, given a chain of length $N$, one can suppose that the
subspaces $\mathcal{A}$ and $\mathcal{B}$ are the left half and the
right half of the chain, respectively. We would like to have the
largest entanglement generation rate. To this end, we focus on the
initial state with zero expectation value of magnetization in the
$z$ direction \cite{Znidaric02,Nanduri01}, i.e. $\vert
\psi_\mathcal{A} \rangle$ and $\vert \psi_\mathcal{B} \rangle$ are
$\vert \uparrow \downarrow \uparrow \cdots \uparrow\downarrow
\rangle$. So, the time-dependent von Neumann entropy can be obtained
via
\begin{equation}\label{Eq28}
S_\mathcal{A} (t) = - Tr \left( \rho_\mathcal{A} (t) ln
(\rho_\mathcal{A} (t)) \right),
\end{equation}
where time-dependent reduced density matrix reads
\begin{equation}\label{Eq29}
\rho_\mathcal{A} (t) = Tr_\mathcal{B} \left( e^{-iHt} \vert \psi (0)
\rangle \langle \psi (0) \vert e^{iHt} \right).
\end{equation}
Note that the above calculation can be done directly by means of
exact diagonalization of the Hamiltonian and regardless of the LIOM
concept. However, respecting to the exponentially growth of Hilbert
space with the increasing number of sites and the necessity of
performing numerical computations for various times, one can not
choose the length of the chain more than $20$ sites due to the lack
of computational resources.\par
In this section, we will indicate that, using the form of the
transformations, we made in the previous section and by exploiting
the concept of LIOMs: (a) the calculations in the MBL phase are
independent of the length of the chain, and (b) with a much smaller
amount of computations resulting in a manageable required memory and
simultaneously with a fairly precise approximation, we are able to
calculate the entanglement generation rate. According to Fig.~\ref{Fig03},
we assume that the Hamiltonian given in Eq.~\eqref{Eq01}
with the blocks of $12$ sites has led to satisfactory results for
Eq.~\eqref{Eq19}, so that we can ignore the off-diagonal terms
of the Hamiltonian. In other words, we approximate the Hamiltonian,
on which the unitary transformations have been acted as follows:
\begin{equation} \label{Eq30}
\mathcal{H} = diag (U^{\dagger} H U).
\end{equation}
Therefore, we can use the following relation to calculate
$\rho_\mathcal{A} (t)$ in the physical-spin basis
\begin{equation} \label{Eq31}
\rho_\mathcal{A} (t) = Tr_\mathcal{B} \left( U U^\dagger e^{-iHt} U
U^\dagger \vert \psi (0) \rangle \langle \psi (0) \vert U U^\dagger
e^{iHt} U U^\dagger \right).
\end{equation}
To put it another way,
\begin{equation}\label{Eq32}
\rho_\mathcal{A} (t) = Tr_\mathcal{B} \left( U e^{-i \mathcal{H} t}
\vert \psi' (0) \rangle \langle \psi'(0) \vert e^{i \mathcal{H} t}
U^\dagger \right),
\end{equation}
where $e^{-i \mathcal{H} t}$ and $\vert \psi' (0) \rangle =
U^\dagger \vert \psi (0) \rangle$ have been written in the LIOM
basis. Note that to obtain the von Neumann entropy in the LIOM
basis, entering $U$ (and $U^\dagger$) in Eq.~\eqref{Eq32} is not
required.\par
Before continuing the calculations, it deserves mentioning that,
generally, $\rho_\mathcal{A} (t)$ is invariant under local
transformations $U_\mathcal{A}$ and $U_\mathcal{B}$, i.e. $S
(\rho_\mathcal{A}) = S (\rho'_\mathcal{A})$ if
\begin{equation} \label{Eq33}
\rho_\mathcal{A} = Tr_\mathcal{B} (\vert \psi'_\mathcal{AB} \rangle
\langle \psi'_\mathcal{AB} \vert) = Tr_\mathcal{B} \left(
U_\mathcal{A}^\dagger \otimes U_\mathcal{B}^\dagger \vert
\psi'_\mathcal{AB} \rangle \langle \psi'_\mathcal{AB} \vert
U_\mathcal{A} \otimes U_\mathcal{B} \right)  = \rho'_\mathcal{A}.
\end{equation}
Another remarkable point is that if we consider diagonal
approximation of the Hamiltonian in the LIOM space, as shown in
Fig.~\ref{Fig05}, we have
\begin{equation}\label{Eq34}
\mathcal{H}=\mathcal{H}_A + \mathcal{H}_{\bar{A}} + \mathcal{H}_B +
\mathcal{H}_{\bar{B}} + \mathcal{H}_{AB} + \mathcal{H}_{\bar{A}A} +
\mathcal{H}_{\bar{B}B},
\end{equation}
in which $\mathcal{A} = A+\bar{A}$ and $\mathcal{B} =
B+\bar{B}$.\par
Hence, all the Hamiltonian terms commute together as they have been
written in the LIOM basis. It is noticeable that
$U_i^{1(2)\mathcal{A}}$ and $U_i^{1(2)\mathcal{B}}$ are the
localized unitaries defined in the subspaces $\mathcal{A}$ and
$\mathcal{B}$, respectively. In other words, $U_i^{1\mathcal{A}}$
are the unitary operators of $i$th block (by counting from
bipartition cut, as the origin, to the left) corresponding to the
subspace $\mathcal{A}$ on the first layer and similarly,
$U_i^{1\mathcal{B}}$ are the unitaries of $i$th block (by counting
from bipartition cut to the right) belong to the subspace
$\mathcal{B}$ on the first layer. In addition, $U_i^{2\mathcal{A}}$
are the unitary operators of block $i$ in the subspace $\mathcal{A}$
and on the second layer (in which $U_{AB}^{2}$ is excluded from
labelling $i$) and $U_i^{2\mathcal{B}}$ are the ones of  $i$th block
in subspace $\mathcal{B}$ on the second layer. Thus, the only
unitary operator between the subspaces $A$ and $B$ is $U_{AB}^2$
illustrated in Fig.~\ref{Fig05}. Now, regarding to
Fig.~\ref{Fig02}, it is obvious that $ \mathcal{U}_1 = \prod_{i}
U_i^{1\mathcal{A}} \prod_i U_i^{1\mathcal{B}} \prod_i
U_i^{2\mathcal{A}} \prod_i U_i^{2\mathcal{B}}$ is a local operator.
So, by substitution of $ \mathcal{U}_1 $ for $U_{\mathcal{A}}
\otimes U_{\mathcal{B}}$ in Eq.~\eqref{Eq33}, Eq.~\eqref{Eq32} leads
to
\begin{equation}\label{Eq35}
\rho_\mathcal{A} (t) = Tr_\mathcal{B} \left( U_{AB}^2 e^{-i
\mathcal{H} t} \vert \psi'(0) \rangle \langle \psi'(0) \vert e^{i
\mathcal{H} t} U_{AB}^{2 \dagger} \right)
\end{equation}
which results in the notion that the difference between the von
Neumann entropy corresponding to the physical-spin basis and that of
the LIOM basis is distinguished by $U_{AB}^2$. As the next step, we
consider $\mathcal{U}_2 =  e^{-i \mathcal{H}_{\bar A} t} e^{-i
\mathcal{H}_{\bar B} t} e^{-i \mathcal{H}_{\bar{A} A} t} e^{-i
\mathcal{H}_{\bar{B} B} t} $ the local operator which commutes with
$U_{AB}^2$. Thus, using the same procedure done for getting
Eq.~\eqref{Eq35}, one may reaches to
\begin{equation}\label{Eq36}
\rho_\mathcal{A} (t) = Tr_\mathcal{B} \left( U_{AB}^2 e^{-i
(\mathcal{H}_A + \mathcal{H}_B + \mathcal{H}_{AB}) t} \vert \psi'
(0) \rangle \langle \psi' (0) \vert e^{i (\mathcal{H}_A +
\mathcal{H}_B + \mathcal{H}_{AB}) t} U_{AB}^{2 \dagger} \right).
\end{equation}
Ultimately, respecting to the commutation relation of $\mathcal{U}_3
= \prod_{i\neq1} U_i^{1\mathcal{A}} \prod_{i\neq1}
U_i^{1\mathcal{B}} \prod_i U_i^{2\mathcal{A}} \prod_i
U_i^{2\mathcal{B}}$ (gray blocks in Fig.~\ref{Fig05}) with $U_{AB}^2
e^{-i (\mathcal{H}_A + \mathcal{H}_B + \mathcal{H}_{AB}) t}$,
the reduced density matrix of Eq.~\eqref{Eq36} can be written as:
\begin{equation}\label{Eq37}
\rho_\mathcal{A} (t) = Tr_\mathcal{B} \left( U_{AB}^2 e^{-i
(\mathcal{H}_A + \mathcal{H}_B + \mathcal{H}_{AB}) t} \vert \psi''
(0) \rangle \langle \psi'' (0) \vert e^{i (\mathcal{H}_A +
\mathcal{H}_B + \mathcal{H}_{AB}) t} U_{AB}^{2 \dagger} \right),
\end{equation}
where $\vert \psi'' (0) \rangle =  U_C^\dagger \vert \psi (0)
\rangle$ and $U_C = U_1^{1\mathcal{A}} U_1^{1\mathcal{B}} U_{AB}^2 =
U_A^1 U_B^1 U_{AB}^2$, in which we show $U_1^{1\mathcal{A}}$ and
$U_1^{1\mathcal{B}}$ as $U_A^1$ and $U_B^1$, respectively.
Consequently, to calculate the entanglement entropy, we just need to
consider the blue blocks exhibited in Fig.~\ref{Fig05} instead of
the chain entirely.\par
That being said, the above calculations results in the following
important achievement which is our main notion of this study:
if the Hamiltonian is diagonalized with an arbitrary approximation under a tensor-network
unitary transformation, to calculate the entanglement, it is sufficient to consider two blocks
next to each other.\par
\begin{figure}[h]
\begin{center}
\includegraphics[width=8cm]{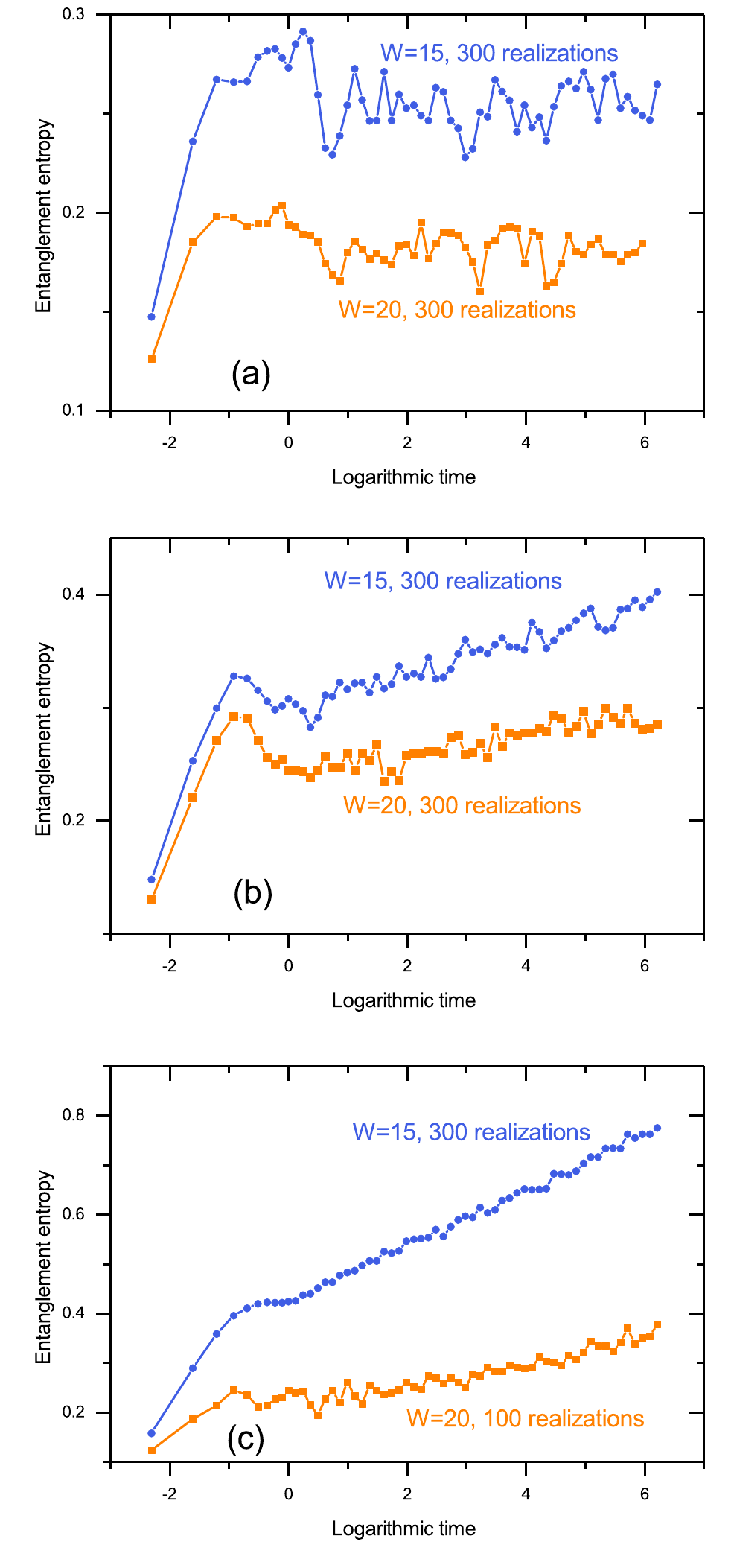}
\end{center}
\vspace*{-3mm} \caption{For different values of the block length,
$l$, the entanglement entropy changes in terms of the logarithm of
time have been computed using the proposed tensor-network
representation: a) for $l=4$, the entanglement entropy quickly
saturates to a constant value, b) for $l=8$, the entanglement
entropy exhibits a weak linear growth at the large times and, c) for
$l=12$, after a relatively long time, the entanglement entropy shows
a linear growth. This behavior is very close to what was reported in
Ref.~\cite{Bardarson01} using the exact diagonalization of the
entire Hamiltonian.    \label{Fig06}}
\end{figure}

To show one of the efficient applications of the present tensor
network method, we have investigated the time evolution of the
entanglement entropy in the system for different block lengths using
Eq.~\ref{Eq37}. For different values of the block length, $l=4,\ 8,$ and $12$,
the entanglement entropy, or von Neumann entropy has been plotted ve the logaritm of time.
The results of the computations are displayed in figuur~\ref{Fig06}.
As is seen from the figure, for a block length of 4, the behavior of the system
is very simple. In the beginning, the system exhibits a rapidly
changing behavior with time. The entropy of the system quickly
reaches its maximum value and then it saturates to a given value. It
has been shown that in an MBL system, the entanglement spreading has
a logarithmic behavior with time, but in this case the amount of
entanglement tends to a constant value so quickly. In fact, since
the length of the block is smaller than the localization length of
the LIOMs, the behavior we expect from an MBL system does not occur
for this case. On the other hand, for a block length of 4, the
available phase space for the entanglement spreading is very small,
so we should not expect logarithmic behavior for this
phenomenon.\par
For a block length of 8, both the localization length of the LIOMs
is of the same order as the block length, and the available phase
space for the entanglement spreading is more considerable.
Consequently, the occurrence of logarithmic behavior for the
entanglement spreading is not unexpected. For this case, such a
logarithmic behavior can be seen for example for W=15 and/or for
W=20 with different slopes.  However, for W=20, the logarithmic
spreading has a very small slope. For a block length of 12, the
logarithmic behavior of the entanglement spreading can be observed.
This observation is consistent with the results reported in Ref.~\cite{Bardarson01} by Bardarson et al.\par
For an acceptable range of the disorder sterngth, $W$, the comparison made in figure~\ref{Fig06} shows
that for the sufficiently large values of the block length, the
local unitary matrices obtained from the approximate tensor-network
method are compatible with those obtained from the exact
diagonalization of the entire Hamiltonian.\par
In the end, it is necessary to note that although the present method has been carried out for the renowned XXZ
model, it can be generalized to the other models for which the LIOMs
can be defined. For example, this scheme can be implemented for a
model of bond disordered XX-spin chain with long-range couplings and
a lattice Schwinger model as well on which we are conducting our
research as a future prospect.\par
\section{Conclusion~\label{Sec04}}
A novel method for constructing the local unitary operators in the
tensor network framework has been proposed, which allows us to
obtain the unitary operator with high accuracy for blocks of very
large length. The effective accuracy of the presented method has
been examined and it has been shown that if the block length is
larger than the localization length of the LIOM, then the entire
Hamiltonian can be diagonalized with a high accuracy by the unitary
operator of the tensor network. Finally, as an application of the
proposed tensor network method, the spreading of the entanglement
entropy of the system with time has been investigated. It is worth
noting that for small block lengths, the entanglement spreading
behavior is not obtained correctly and the entanglement entropy
tends quickly to a constant value. But with the increase of the
block length, the entanglement spreading exhibits such a behavior
that is a characteristic of the MBL systems. When the block length
increases to~$12$, for for the disorder strengthes of~$15$ and~$20$,
the behavior of the system is in very good consistent with the
results of Bardarson~et~al, obtained using the exact diagonalization
of the Hamiltonian and reported in Ref.~\cite{Bardarson01}. Our
study, firstly shows that the tensor network method with length~$12$
can be used effectively for a large range of the disorder strength
W,  and secondly, the entanglement propagation reaches the limits of
the block length of~$12$ for large times. In other words, the
logarithmic spreading of the entanglement is confined in a region of
length~$12$ for large times.
\section*{Data availability statement}
 The datasets generated and analyzed during the current study
 are available from the corresponding author on reasonable request.

\end{document}